\documentclass[journal]{IEEEtran}
\IEEEoverridecommandlockouts
\usepackage{cite}
\usepackage{amsmath,amssymb,amsfonts}
\usepackage{algorithmic}
\usepackage{graphicx}
\usepackage{textcomp}
\usepackage{xcolor}
\def\BibTeX{{\rm B\kern-.05em{\sc i\kern-.025em b}\kern-.08em
    T\kern-.1667em\lower.7ex\hbox{E}\kern-.125emX}}

\usepackage{flushend}
\usepackage[retain-explicit-plus=true]{siunitx}
\DeclareSIUnit \year { a }
\DeclareSIUnit \pu { p.u. }
\DeclareSIUnit \voltampere { VA }
\DeclareSIUnit \kWh { kWh }
\usepackage[nolist]{acronym}
\usepackage[hidelinks]{hyperref}
\usepackage{cleveref}
\crefformat{equation}{(#2#1#3)}
\usepackage{float}
\usepackage{pgfplots}
\pgfplotsset{compat=1.17}
\usepackage{pgfplotstable}
\usepackage{multirow,tabu}
\usepackage{booktabs} 
\usepackage{verbatim} 

\usepackage{caption}
\usepackage{subcaption}
\usepackage{stfloats} 
\usepackage{lipsum}
\usepackage{array} 
\usepackage{bm}

\newcommand{\trafovoltages}[2]{\qtylist[list-units = single,list-pair-separator = {/}]{#1;#2}{\kV}}

\begin{document}
\begin{acronym}
    \acro{OSM}{OpenStreetMap}
    \acro{DSO}{distribution system operator}
    \acro{RNM}{Reference Network Model}
    \acro{RMS}{root mean square}
    \acro{EMT}{electro-magnetic transient}
    \acro{HV}{high-voltage}
    \acro{MV}{medium-voltage}
    \acro{LV}{low-voltage}
    \acro{TSP}{travelling salesman problem}
    \acro{PV}{photovoltaics}
    \acro{MILP}{mixed integer linear program}
    \acro{DG}{distributed generation}
    \acro{EM}{electricity meter}
\end{acronym}

\title{Open Data-Driven Automation of Residential Distribution Grid Modeling with Minimal Data Requirements}

\makeatletter
\newcommand{\linebreakand}{%
  \end{@IEEEauthorhalign}
  \hfill\mbox{}\par
  \mbox{}\hfill\begin{@IEEEauthorhalign}
}
\makeatother

\author{
Moritz Weber,
Luc Janecke,
H{\"u}seyin K. \c{C}akmak,
Veit Hagenmeyer,~\IEEEmembership{Member,~IEEE}%
\thanks{This work was conducted within the framework of the Helmholtz Program Energy System Design (ESD). The authors gratefully acknowledge funding by the German Federal Ministry of Education and Research (BMBF) within the Kopernikus Project ENSURE ‘New ENergy grid StructURes for the German Energiewende’.}
\thanks{M. Weber, L. Janecke, H. K. \c{C}akmak, and V. Hagenmeyer are with the Institute for Automation and
	Applied Informatics, Karlsruhe Institute of Technology, 76344 Eggenstein-Leopoldshafen,
	Germany (e-mail: moritz.weber@kit.edu).}
}

\maketitle

\begin{abstract}
In the present paper, we introduce a new method for the automated generation of residential distribution grid models based on novel building load estimation methods and a two-stage optimization for the generation of the 20 kV and 400 V grid topologies.
Using the introduced load estimation methods, various open or proprietary data sources can be utilized to estimate the load of residential buildings.
These data sources include available building footprints from OpenStreetMap, 3D building data from OSM Buildings, and the number of electricity meters per address provided by the respective \ac{DSO}. 

For the evaluation of the introduced methods, we compare the resulting grid models by utilizing different available data sources for a specific suburban residential area and the real grid topology provided by the \ac{DSO}.
This evaluation yields two key findings:
First, the automated 20 kV network generation methodology works well when compared to the real network.
Second, the utilization of public 3D building data for load estimation significantly increases the resulting model accuracy compared to 2D data and enables results similar to models based on \ac{DSO}-supplied meter data.
This substantially reduces the dependence on such normally proprietary data.

\end{abstract}

\begin{IEEEkeywords}
model generation, open data, open-source software, optimization, distribution grid, grid capacity analysis
\end{IEEEkeywords}

\section{Introduction}

The global path to carbon-neutral energy generation leads to more renewable energy sources, such as wind parks and photovoltaic systems, in the electricity grid \cite{Zinaman.2018, Miller.2015}.
These renewable energy sources are often distributed throughout the grid, instead of the traditionally centralized energy generation by fossil fuels or nuclear power.
As such, these decentralized energy sources are often located in the lower grid levels, which leads to an increased generation in the distribution grid level, especially through rooftop photovoltaics.
This requires detailed simulations of all grid levels, from \ac{HV} to \ac{LV}.
Simulations of the \ac{MV} and \ac{LV} grids are particularly important since those grids were not designed to incorporate large quantities of generation originally.
Furthermore, these simulations are crucial to develop local and district-based solutions for challenges caused by the transition to carbon-neutral energy generation.
However, the required grid data and models often are not readily available due to data privacy concerns, necessitating often labor-intensive modeling processes before the actual simulations can be performed.
Thus, the goal of the present work is to further develop methods for the automated creation of grid models, building on our previous work \cite{Cakmak.2022.ISGT}, using readily available open data sources.
While the goal is not the exact recreation of existing grids, the generated models aim to be realistic enough that they could describe the real grid.
These grid models are intended to support various use cases, e.g., machine learning applications and realistic simulations on a wide range of simulation software, such as load flow calculations, quasi dynamic simulations, \ac{RMS} and \ac{EMT} simulations.

The main contribution of the present paper is a new two-stage optimization method for the automated generation of \ac{LV} grids relying solely on openly available data sources, such as \ac{OSM} \cite{OSMContrib.2004} and OSM Buildings \cite{OSMBuildings.2022} that further enables the automated model generation of the \qty{20}{\kV} medium-voltage grid.
This significantly reduces the data requirements compared to previous approaches \cite{Cakmak.2022.ISGT,MateoDomingo.2011.full,InstituteforEnergyandTransportJointResearchCentre2016DistributionSystem,Grzanic.2019,Mateo.2020, Krishnan.2020, Palmintier.2021,Abhilash.2021,Medjroubi.2017,Amme.2018,Klabunde.2022}.
Furthermore, we present a comprehensive comparison of various data sources -- open and proprietary -- for the automated generation of distribution grid models.

The remainder of this paper is structured as follows:
In \Cref{sec:related-work}, we provide an overview of current related work.
In \Cref{sec:method}, we describe the methodology behind our proposed model generation method, including the building load estimation with various data sources, and our optimization-based transformer placement method.
We then evaluate the proposed method and compare the results using the different data sources in \Cref{sec:eval} before discussing the results in \Cref{sec:discussion}.
Finally, we conclude with our findings and give a brief outlook on our future work in \Cref{sec:conclusion}.

\section{Related Work}
\label{sec:related-work}

Since the demand for distribution grid models is apparent for many use cases centering around grid simulations and studies, the data-driven generation of such models is a well-researched topic.

The process for creating so-called \acp{RNM} described in \cite{MateoDomingo.2011.full} marks a fundamental work in the field of automated power grid modeling.
This approach utilizes data on the location and demand of customers, locations and capacity of \ac{DG} and transmission substations, and economic and technical parameters to generate European-style power grid models from the \ac{HV} level down to individual \ac{LV} customers.
The grid models are created by a heuristic branch-exchange method to minimize the cost of the grid using a pre-defined catalog of standard equipment.
Building on this, \cite{InstituteforEnergyandTransportJointResearchCentre2016DistributionSystem} creates and publishes several models of selected test areas in the \textsc{Matpower} \cite{Matpower.2011} format and compares key indicators of these models with real-world data provided by European \acp{DSO}.
A further development of this process is described in \cite{Grzanic.2019}, which focuses on the development of an online tool for the automated model generation from \ac{OSM} data, area parameters (e.g. consumer density, power factor, and \ac{MV}/\ac{LV} transformer locations), and \ac{DSO} indicators.

More enhancements to \cite{MateoDomingo.2011.full} are described in \cite{Mateo.2020, Krishnan.2020, Palmintier.2021} with the adaption to U.S.-style distribution grids and an appropriate validation approach for this grid type.
The approach introduced in \cite{Mateo.2020} is capable of creating complex U.S.-style power grids with their typical single phase connections and voltage regulators on the \ac{LV} level.
While the previously mentioned methods expect the location and demand of customers as a direct input, \cite{Mateo.2020} describes a method for estimating this demand based on land use data (usually from commercial vendors) and a library of reference buildings.
Considering the complex nature of U.S.-style distribution grids, this approach creates detailed models in the OpenDSS \cite{OpenDSS.2017} and CYME format.

Other approaches, such as  \cite{Abhilash.2021} focus on the German power grid.
This approach utilizes \ac{OSM} data and the known total number of \ac{LV} networks in Germany to generate a total of \num{500000} \ac{LV} distribution grid topologies.
The generated topologies are validated on a statistical basis with real grid data, such as the number of nodes and edges per \ac{LV} grid and the total line length.
Other approaches in the same research context focus on the generation of transmission grid \cite{Medjroubi.2017} and \ac{MV} grid \cite{Amme.2018} models instead of the lower voltage levels.
The method described in \cite{Klabunde.2022} is focused on regional distribution grid structures as found in Germany and utilizes nine different data sources to classify buildings for load estimation and for generating the \ac{MV} and \ac{LV} grid topologies.
The method generates \textsc{Matpower} \cite{Matpower.2011} models and is validated on a high-level basis against real networks.

Despite all of these approaches, the research in this domain is still open in a few key areas:
Most of the methods available in the literature utilize some kind of proprietary data or very specific knowledge that prohibits a wider applicability.
Furthermore, to the best of our knowledge, there has not been a comparison of models generated using different available data sources.
Lastly, many available methods create models in very basic formats, such as \textsc{Matpower}.
In the present work, we address those three areas by presenting a method that requires minimal data input and generates versatile PowerFactory models.
Furthermore, we compare the generated models under different input data circumstances.

\section{Methodology}
\label{sec:method}

In this section, we first describe the building load estimation for low-voltage grids before describing the two optimization stages of the grid model generation as shown in \Cref{fig:overview}.
The load estimation process uses various open or proprietary data sources to estimate the load per building that is a required input of the two-stage grid layout optimization.
In the first stage of this optimization, the \qty{20}{\kV} grid topology with its \trafovoltages{20}{0.4} substations is generated using a k-means clustering approach for the substation placement and a \ac{TSP} optimization for the line routing between the stations.
In the second stage, the underlying \qty{400}{\volt} grid is generated by solving a variation of the minimum cost flow linear optimization problem.
Since the methodology described in this work focuses on areas dominated by underground cables instead of overhead lines, whose routes might be available in map data, we make the common assumption that cables are laid out along roads and paths.

\begin{figure}[t]
    \includegraphics[width=\linewidth]{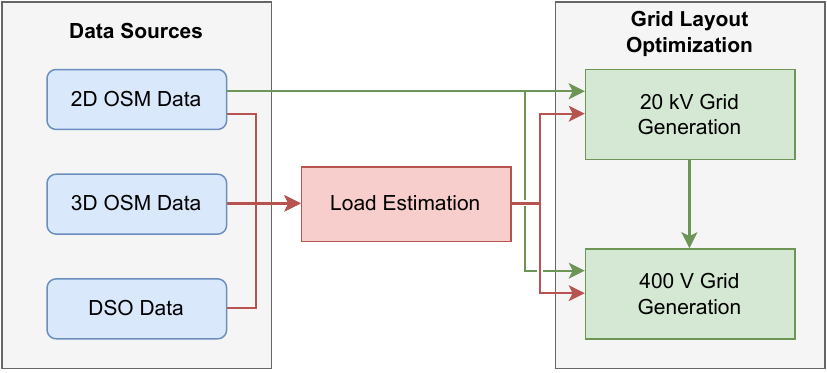}
    \caption{Two-stage optimization method for automated distribution grid generation with load estimation based on various alternative data sources.}
    \label{fig:overview}
\end{figure}

\subsection{Building Load Estimation Based on Variable Data Sources}
\label{sec:method:load}

Since most existing distribution grids evolved over many years and traditionally only loads were considered in the planning of low-voltage distribution grids, we neglect \ac{PV} generation in our methodology for estimating the required sizes of equipment, such as transformers and cables.
The impact of renewables in the low-voltage grid is rather part of studies that can be carried out with the generated grid models, see e.g. \cite{Cakmak.2022}.
Thus, the automatically generated network topologies mainly depend on the assumed loads for the identified buildings in the modeled area, since the network topology is generated with an optimization algorithm based on load data.

For the building load estimation, we consider three different open and proprietary data sources.
These data sources are \ac{OSM} data $(O_{\mathit{2D}})$, 3D OSM Buildings $(O_{\mathit{3D}})$ and finally the number and location of electricity meters provided by a local distribution system operator $(EM)$.
The load estimation methods described in the following are, however, used for residential buildings only.
Non-residential buildings are identified via tags included in \ac{OSM} data and require a special treatment.

In order to estimate the load of a residential building $i$, we utilize the household standard load profile $\text{H}0$ \cite{Meier.1999} that is multiplied by an estimated yearly energy consumption $E_i$:
\begin{equation}\label{eq:P_it}
        P_i(t) = \text{H}0(t) \cdot E_i
\end{equation}
The estimation of the yearly consumption of a building starts with estimating the energy consumption for a single residential unit $E_{U}$,
\begin{equation}\label{eq:e2d_ru}
    E_{U}(A) = nR \cdot E_R + A \cdot E_A + nLA \cdot E_{LA},
\end{equation}
where $nR$ is the average the number of residents in a household according to \cite{Zensus.2011.dt}, $A$ is the floor area of the residential unit, and $nLA$ is the statistical number of large electrical appliances per household for the selected test area \cite{bde.2004}.
Specific values used for the application of this method can be found in \Cref{tab:specific_values}.

The number of residential units in a building $i$, if only the floor area is considered, is estimated as
\begin{equation}\label{eq:nH_2D}
    nU^{\mathit{2D}}_i = 
        \begin{cases}
        A_i / A_U, & \text{if } A_i / A_U \leq 1 \\ 
        1, & \text{else},
    \end{cases}
\end{equation}
where $A_U$ is the median floor area of a residential unit.
This median value $A_U$ is determined over all floor areas of the buildings $A_i$ in the target region which geographically defines an area:
\begin{equation}\label{eq:mA}
    A_U = median\left(A_i\right).
\end{equation}
To model the energy consumption of individual buildings $E_{i}$, which is based on the consumption per unit $E_U$, the base areas of the buildings are decisive and may vary greatly depending on the data basis. 

\subsubsection{\ac{OSM} data to determine the base area of buildings}
Using the building floor area, the energy consumption per building $E^{\mathit{2D}}_{i}$ is obtained by multiplying $E_U$ with the number of residential units $nU^{\mathit{2D}}_i$, see \cref{eq:nH_2D}, and a scaling factor $S_U$,
\begin{equation}\label{eq:P_2D}
    E^{\mathit{2D}}_{i} = nU^{\mathit{2D}}_i \cdot S_U \cdot E_{U}(A_U).
\end{equation}
The scaling factor $S_U$ can be approximated by the average number of stories per building and can be adjusted downward to accommodate for a larger proportion of single-family homes.

\subsubsection{\ac{OSM} data and height information from OSM Buildings}
In this estimation, the building height $H_i$ is considered to approximate the number of floors of a building using data provided by OSM Buildings \cite{OSMBuildings.2022}.
According to \cite{BW_GebEnergie_Gesetz.2020} the floor height is $2.5-3m$.
In this range, an average floor height $h_f$ is chosen for the further calculation.
The corresponding energy consumption $E^{\mathit{3D}}_{i}$ is calculated as
\begin{equation}\label{eq:P_3D}
    E^{\mathit{3D}}_{i} = nU^{\mathit{2D}}_i \cdot H_i/h_f \cdot E_{U}(A_U),
\end{equation}
where the additional factor $H_i/h_f$ replaces the previously introduced scaling factor and is individual for each building $i$.

\subsubsection{Electricity meter data supplied by public utilities}
The information about the number of \acp{EM} for all buildings $nU^{EM}_i$ provided by the public utilities (DSO) is used for the direct calculation of the year-round load data.
In this data, an electricity meter for general electricity is given for buildings with multiple households.
This additional meter is considered in this context with the fictitious electricity consumption of \num{0.1} households.
The load for a building can be estimated as follows,
\begin{equation}\label{eq:E_EM}
    E^{EM}_i = \begin{cases}
        nU^{EM}_i \cdot E_{U}\left(A_{U,i}\right), & \text{if } nU^{EM}_i < 3 \\ 
        \left(nU^{EM}_i - 0.9\right) \cdot E_{U}\left(A_{U,i}\right), & \text{else},
    \end{cases}
\end{equation}
where $A_{U,i}$ is the average floor area of a residential unit in the building $A_i$.
$A_{U,i}$ is calculated as
\begin{equation}\label{eq:A_Ui}
   A_{U,i} = A_i \cdot H_i/h_f~/~nU^{EM}_i.
\end{equation}
The generated load time series,  as described in \cref{eq:P_it}, can then be utilized to perform quasi-dynamic simulations to evaluate the models under a wide range of conditions.

\subsection{Stage 1: Automated Generation of the 20kV Grid}
\label{sec:method:stage1}

In this section, we introduce the first stage of the optimization-based grid generation that handles the creation of a \qty{20}{\kV} distribution grid.
This stage is based on map data and the load estimation introduced in the previous section, and consists of two consecutive steps:
First, the estimation of the number and locations of the \trafovoltages{20}{0.4} substations and second, the grid topology generation connecting these substations.

\subsubsection{Calculation of the Number and Locations of Substations}
\label{sec:method:substations}

In order to estimate the adequate number of transformers for the grid, the results of the load estimation are used.
This estimation is based on the peak load share per building $P_{peak,i}$ that is calculated as
\begin{equation}\label{eq:p_peak}
    P_{peak,i} = P_{peak,U} \cdot nU_i,
\end{equation}
where $nU_i$ is the estimated number of residential units of building $i$, which depends on the utilized data source.
In accordance with \cite{Agricola.2012}, each household accounts for a peak load share of \qty{2}{\kW}.

Assuming a power factor $\lambda$ for the household loads and a loading of at most $L_T$ for each of the \trafovoltages{20}{0.4} substations with a rating of $R_T$, the needed number of transformers $nT$ is determined with the ceiling function applied to the quotient of the total amount of power $P$ that is consumed by each of the $N$ buildings and the adjusted transformer rating.
This leads to the estimation of the number of transformers $nT$ as
\begin{equation}\label{eq:nr_trafos}
    nT = \left \lceil \frac{\sum_{i=1}^N P_{peak,i}}{\lambda \cdot R_T \cdot L_T} \right \rceil.
\end{equation}
In order to then compute the locations of the stations, a \textit{k-means clustering} algorithm is used \cite{Lloyd.1982}.
Each of the $nT$ returned clusters is based on each building's geographical coordinates, uses the case-specific number of households per building $nU_i$ as the weight for the computation, and has a cluster center that is used for the transformer location. 
Once this is done, a graph representation of the street layout is obtained, that includes the closest \ac{HV}/\ac{MV} substation and the area where the grid is to be generated.
Then the calculated station positions are added as nodes to this graph structure in the same way that each building is appended.

\begin{table}
    \centering
    \caption{Specific values used for the load and transformer estimation.}
    \label{tab:specific_values}
    \resizebox{\columnwidth}{!}{
    \begin{tabular}{llrr}
        \toprule
        \textbf{Symbol} & \textbf{Description} & \textbf{Value}  \\
        \midrule
        $nR$ & Average number of residents per unit \cite{Zensus.2011.dt} & 1.7 \\
        $E_R$ & Consumption per resident \cite{bde.2004} & \qty{200}{\kWh}\\
        $E_A$ & Area-dependent consumption \cite{bde.2004} & \qty{9}{\kWh\per\meter\squared}\\
        $nLA$ & Number of large appliances \cite{bde.2004} & \num{8.4} \\
        $E_{LA}$ & Consumption per large appliance \cite{bde.2004} & \qty{200}{\kWh}\\
        \midrule
        $h_f$ & Average floor height \cite{BW_GebEnergie_Gesetz.2020} & \qty{2.6}{\meter} \\
        \midrule
        $P_{peak,U}$ & Simultaneous peak load per residential unit \cite{dena.2012} & \qty{2}{\kilo\watt} \\
        \midrule
        $\lambda$ & Power factor & \num{0.95} \\
        $R_T$ & Transformer rating & \qty{630}{\kilo\voltampere} \\
        $L_T$ & Target transformer loading & \qty{50}{\percent} \\
        \bottomrule
    \end{tabular}
    }
\end{table}

\subsubsection{Optimization-based 20kV Grid Topology Generation}
An optimization approach is then used to determine the associated network topology for the identified \qty{20}{\kV} stations. 
The problem is modeled as a travelling salesman problem \cite{Stein.2022}, with the starting and ending point being the \ac{HV}/\ac{MV} substation.
The places to visit are the \qty{20}{\kV} stations and the distances are the shortest paths on the corresponding graph that is weighted with the actual, geographical length of each road.
This problem is approximated using the Christofides algorithm \cite{Christofides.1976} and can be formalized as follows:

\begin{equation}
\label{eq:tsp1}
\text{min} \sum_{(i,j) \in Paths} l_{i,j} \cdot x_{i,j}
\end{equation}

\begin{equation}
\label{eq:tsp2}
x_{i,j} = x_{j,i} \quad \forall (i,j) \in Paths
\end{equation}

\begin{equation}
\label{eq:tsp3} \sum_{(i,j) \in Paths} x_{i,j} = 2 \quad \forall i \in Stations
\end{equation}

\begin{equation}
\label{eq:tsp4}
\sum_{(i \neq j) \in S} x_{i,j}  \leq  \left| S \right| - 1 \quad \forall S \subset Stations
\end{equation}

The \ac{MV}/\ac{HV} substation together with the \qty{20}{\kV} stations form the set of nodes, $Stations$, while $Paths$ is the set of the shortest paths $(i,j)$ between each of the nodes in $Stations$.
Furthermore, $l_{i,j}$ is the length of the shortest path between node $i$ and node $j$.
The decision variable $x_{i,j}$ equals 1 if the path between node $i$ and node $j$ is part of the solution, and 0 otherwise. 
\cref{eq:tsp2} ensures symmetry while \cref{eq:tsp3} makes sure that each station is connected to two other ones.
The last constraint \cref{eq:tsp4} ensures that the solution is indeed a single \qty{20}{\kV} ring and not just a union of other, smaller rings.

\subsection{Stage 2: Automated Generation of the Low-Voltage Grid}
\label{sec:method:stage2}

In our previous approach \cite{Cakmak.2022.ISGT}, residential \ac{LV} grid models are derived from the street layouts available in \ac{OSM} as a possible cable route. 
In the present paper, a Python application converts these layouts to a graph representation and adds the buildings of the residential area as found in \ac{OSM}.
For each of these buildings, load data is computed using one of the datasets described in Section \ref{sec:method:load}. 
In order to obtain the grid topology, a variation of the minimum cost flow optimization problem modifies the graph to comply with electrical low-voltage grid topology standards.
This optimization is carried out with specified \qty{20}{\kV} substation locations. 
The optimization problem is formulated as a \ac{MILP} with binary decision variables and is stated in the following equations: 
%
\begin{equation}
\label{eq:nfp1}
\text{min} \sum_{(i,j) \in E} cost_{i,j} \cdot install_{i,j}
\end{equation}
%
\begin{equation}
\label{eq:nfp2}
\sum_{j:(i,j) \in E} flow_{i,j} \quad - \sum_{j:(j,i)\in E} flow_{j,i} = residual_i \quad \forall i \in V
\end{equation}
%
\begin{equation}
\label{eq:nfp3}
\sum_{i:(i,j) \in E} install_{i,j}  \le 1 \quad \forall j \in V
\end{equation}
%
\begin{equation}
\label{eq:nfp4}
0 \le flow_{i,j} \le cap^{max}_{i,j} \cdot install_{i,j} \quad \forall (i,j) \in E
\end{equation}
Generally speaking, the program decides which of the edges of the graph will be used as cables for power delivery by choosing the cheapest way to supply all demand.
Therefore, the objective function \eqref{eq:nfp1} minimizes the cost $cost_{i,j}$ to install a cable from node $i$ to node $j$ that is proportional to the potential cable route length. 
A binary decision variable $install_{i,j}$ is introduced in order to only account for potential routes hat are actually used or, in other words, where a cable is installed and the route used to supply buildings with power:
It is 1 if the cable is installed and 0 in all other cases.
The first flow conservation constraints \eqref{eq:nfp2} ensure that for all nodes $i$ that are not a source, the power flowing in $flow_{j,i}$ equals the consumption $residual_i$ of the node itself or is flowing out again. 
For sources that represent the secondary substations, the power flowing out equals the externally provided power that is also denoted $residual_i$. 
A positive $residual_i$ indicates an external source, while a negative one indicates consumption.
For this consumption, possible PV injection is not considered in the optimization out of the aforementioned historical reasons, only the load data computed in the previous steps is used.
The second constraints \eqref{eq:nfp3} ensure radiality in the obtained grid, as \ac{LV} distribution grids are usually operated in a radial mode.
Hence, only one cable can be used to supply a node with power.
Lastly, the constraints \eqref{eq:nfp4} make sure that a certain loading limit $cap^{max}_{i,j}$ for the cables is not exceeded and that energy can only flow over a cable that is also installed.
The algorithm repeats this optimization while lowering the maximum cable loading limit permitted in the optimization for a cable that is installed alongside a public way, in order to obtain a topology that uses all the secondary substations equally. 
In this context, this limit is also referred to as the cable capacity.
In this improved version of the cable capacity estimation (CCE), the approach for lowering this capacity changed from the Newton Bisection method (NB-CCE) described in \cite{Cakmak.2022.ISGT} to the proposed Inverse Proportional approximation (IP-CCE):
\begin{equation}
\label{eq:cap}
cap^{max}_{i+1} = \mathrm{round}\left(cap^{max}_{i}\left(1 - \frac{1}{i + N}\right)\right),
\end{equation}
where $cap^{max}_{i}$ describes the maximum cable capacity in the $i$-th iteration of the optimization and $N$ is an adjustable parameter.
Once the capacity restriction of an iteration makes the problem infeasible, the capacity of the last feasible iteration is selected and the results of that optimization are used for the rest of the workflow.

This new approximation allows for fast progress at the start of the optimization method and finer steps towards the end. 
This ensures that the found final capacity is actually close to the theoretical optimum that is the lowest integer capacity value with which the model remains feasible and that the obtained solution is still balanced regarding secondary substation use. 
Nonetheless, the application still finds good results much faster because, in contrast to the Newton bisection method, very time-consuming optimization steps around the theoretical optimum are avoided.
Experiments with this formula have shown that $N = 4$ yields an especially favorable balance between runtime and quality of the results.

\subsection{PowerFactory Model Generation}
\label{sec:method:powerfactory}

\begin{figure}[t]
    \includegraphics[width=\linewidth]{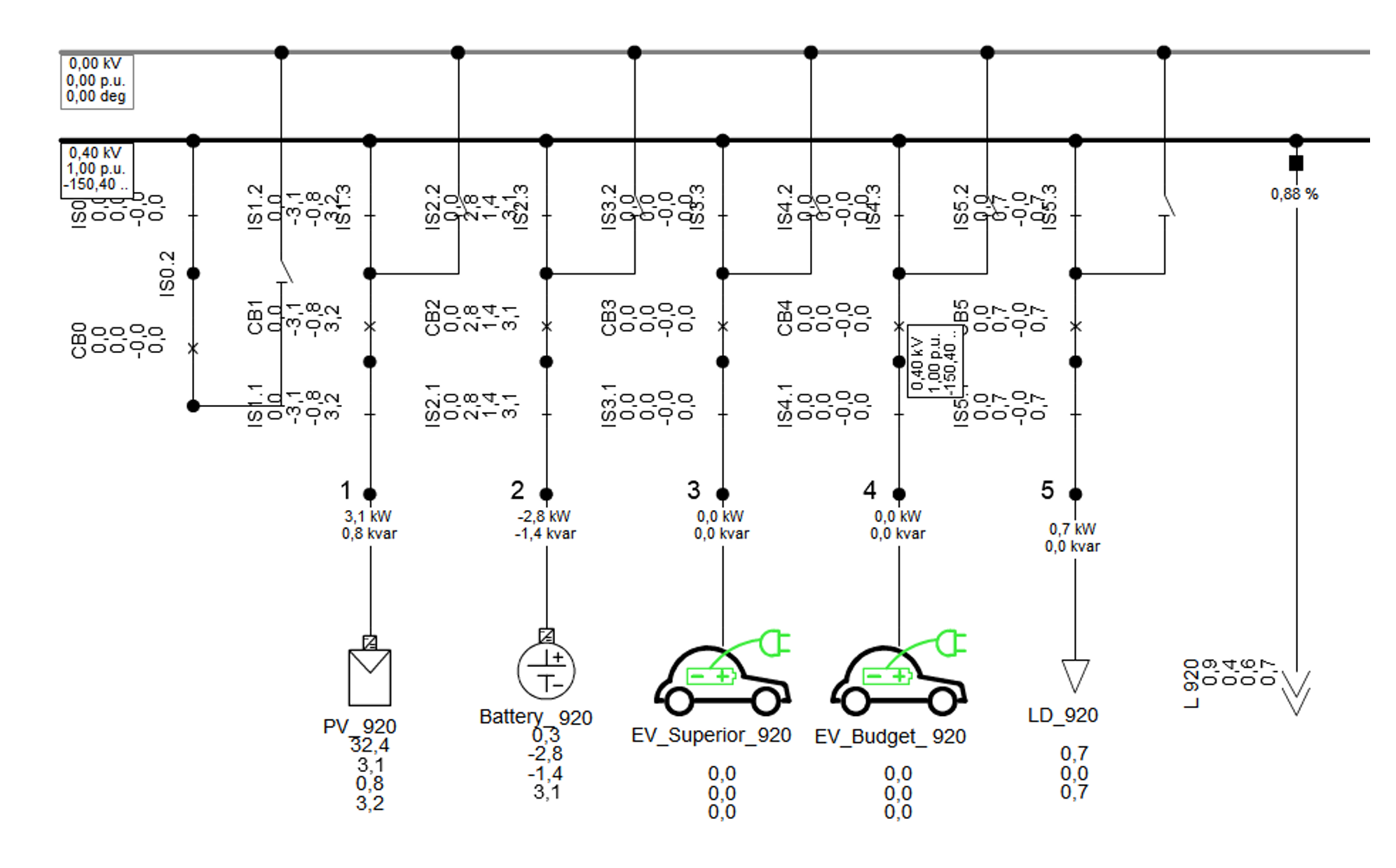}
    \caption{Detailed view of the automatically generated components in each house model. Note that for the analysis in this study, only the residential load is considered. However, this approach offers the flexibility required to study various potential future scenarios.}
    \label{fig:house_model}
\end{figure}

In the last step, the obtained grid data structure is converted to a DIgSILENT PowerFactory model.
Since system parameters such as line impedances cannot be estimated from the utilized data sources, this approach uses a library of system components that are widely used in the target area.
This library can, of course, be adapted to reflect component choices of the local \ac{DSO} if these are known.
This also includes the automated creation of grid diagrams and an initial load flow calculation.
For each individual building, a busbar system as shown in \Cref{fig:house_model} can be created.
Although the present paper only describes the methodology to estimate the loads for these house models, this approach offers the flexibility required for the analysis of future scenarios with various \ac{PV}, battery and electric vehicle settings.
This Python tool chain is fully automated, relies solely on a good OSM data coverage of the target area, and the obtained grids can be readily used for power grid analysis studies.
However, due to varying grid design approaches around the world, its usage is limited to regions with European-style distribution grids.

\begin{figure*}[ht]
\begin{subfigure}[t]{0.32\linewidth}
    \includegraphics[keepaspectratio, height=4cm]{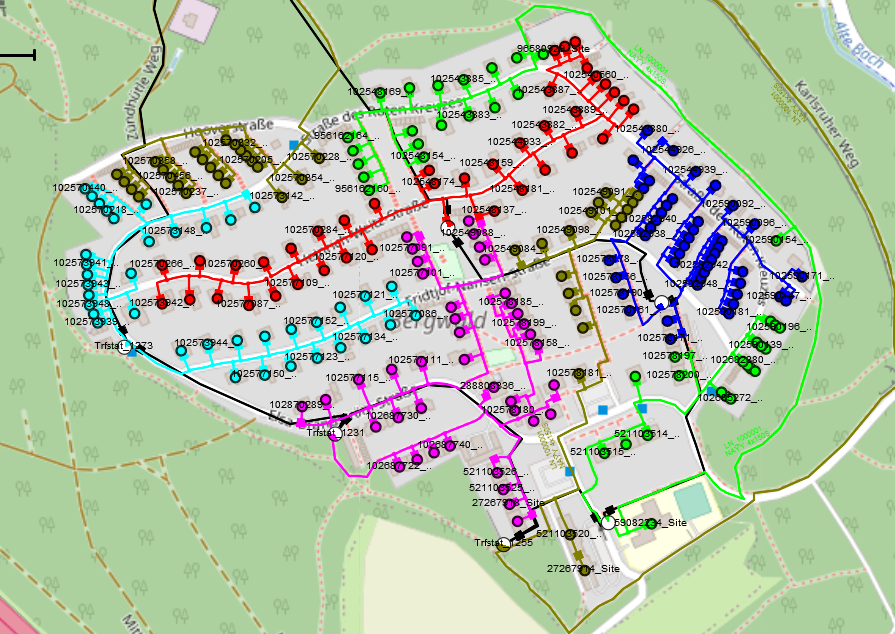}
    \caption{Case $(O_{\mathit{2D}}, T_{K})$: Grid model using the OSM dataset for given number and location of transformers.}
    \label{fig:o2d-tk}
\end{subfigure}
    \hfill
\begin{subfigure}[t]{0.32\linewidth}
    \includegraphics[keepaspectratio, height=4cm]{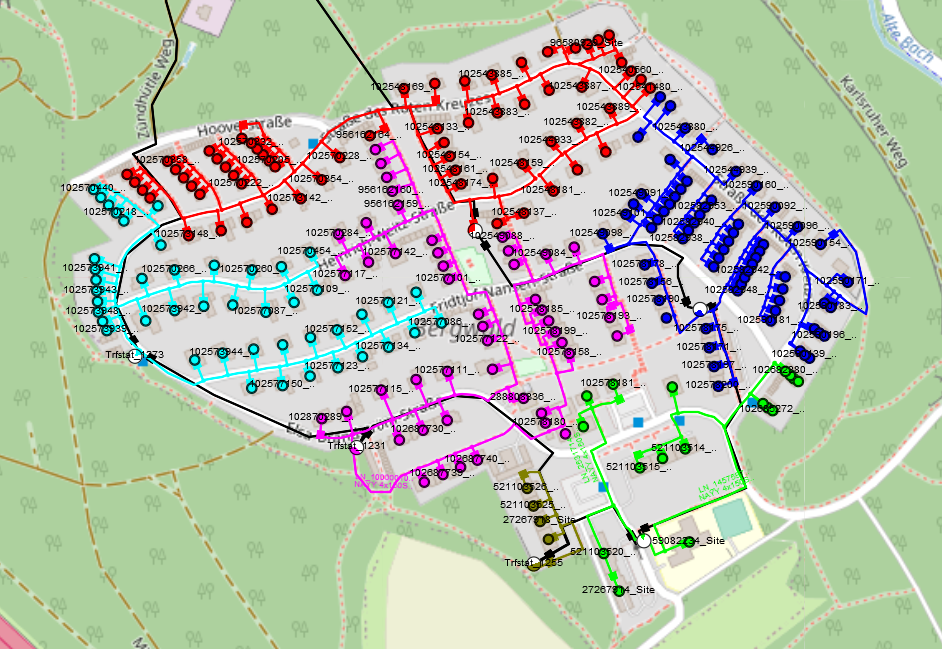}
    \caption{Case $(O_{\mathit{3D}}, T_{K})$: Grid model using the OSM Buildings dataset with included height data of buildings for given number and location of transformers.}
    \label{fig:o3d-tk}
\end{subfigure}
    \hfill
\begin{subfigure}[t]{0.32\linewidth}
    \includegraphics[keepaspectratio, height=4cm]{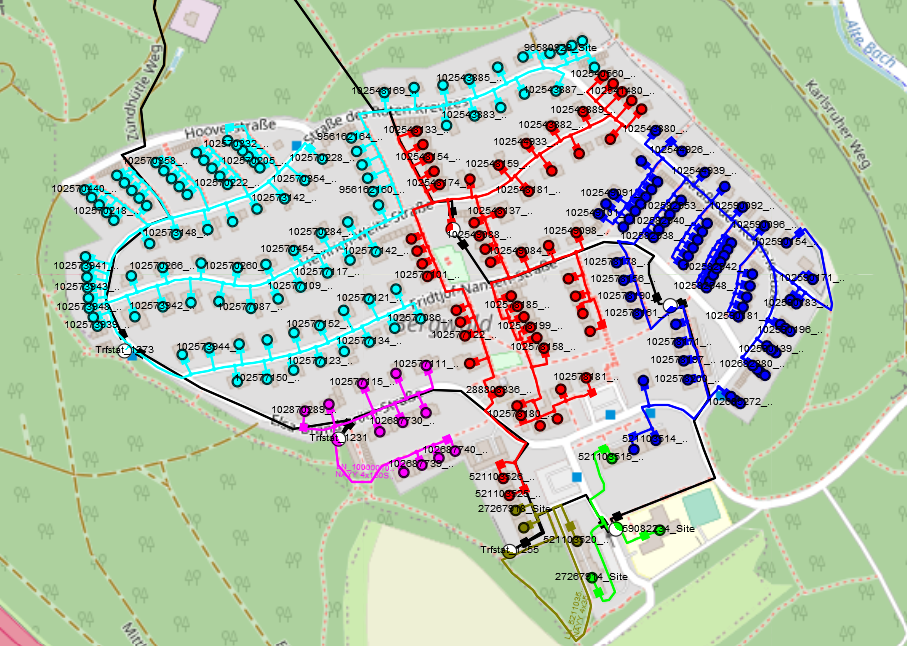}
    \caption{Case $(EM, T_{K})$: Grid model using the \ac{DSO} data on the number of electricity meters at each building for a given number and location of transformers.}    
    \label{fig:em-tk}
\end{subfigure}

\caption{Distribution grid modeling with \textit{a priori known} number and location of substations using different data sources for the load modeling comprising the combinations $(X, T_{K})$. Note that each building is modelled as a subsystem with a \ac{PV} model, residential load, cable and as a preparation for future applications with a battery. The color of the nodes and lines indicates the affiliation to a specific transformer, with a consistent coloring between the three variants.}
\label{fig:1all} 
\end{figure*}
\begin{figure*}[t!]
\begin{subfigure}[t]{0.32\linewidth}
    \includegraphics[keepaspectratio, height=4cm]{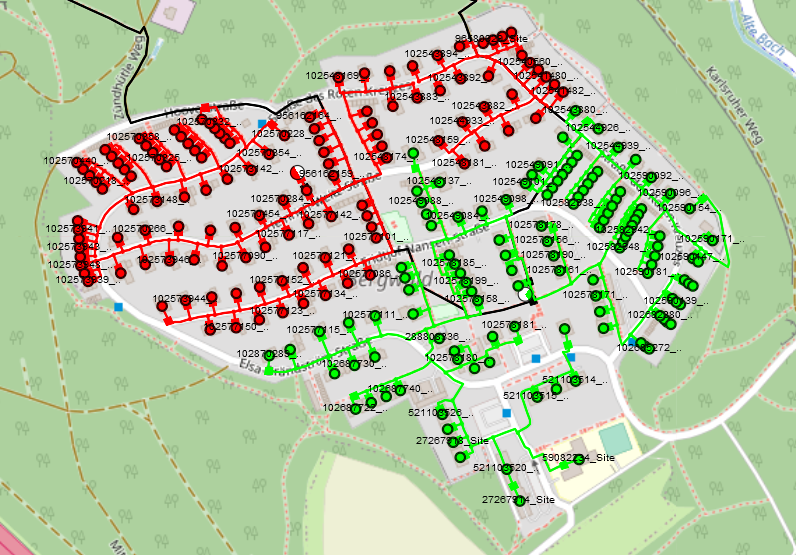}
    \caption{Case $(O_{\mathit{2D}}, T_{C})$: Grid model using the \ac{OSM} dataset for a calculated number and location of transformers.}
    \label{fig:o2d-tc}
\end{subfigure}
    \hfill
\begin{subfigure}[t]{0.32\linewidth}
    \includegraphics[keepaspectratio, height=4cm]{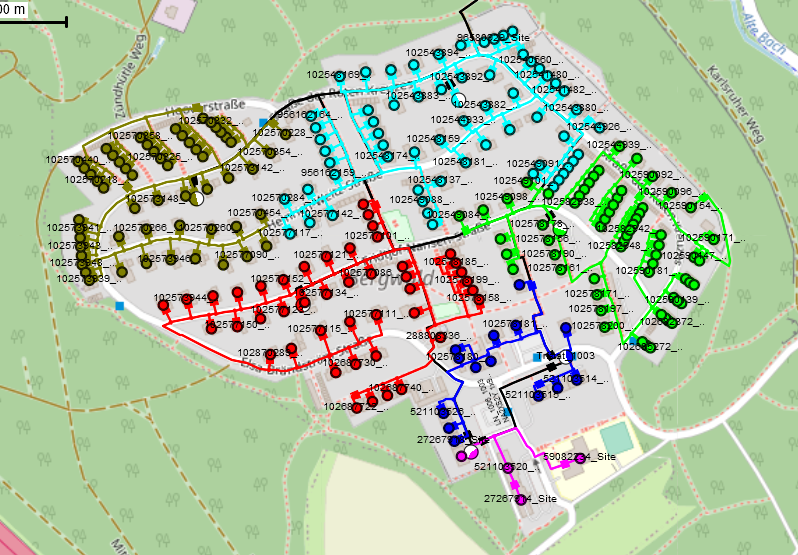}
    \caption{Case $(O_{\mathit{3D}}, T_{C})$: Grid model using the \ac{OSM} Buildings dataset with included height data of buildings. The number and location of transformers are calculated.} 
    \label{fig:o3d-tc}
\end{subfigure}
    \hfill
\begin{subfigure}[t]{0.32\linewidth}
    \includegraphics[keepaspectratio, height=4cm]{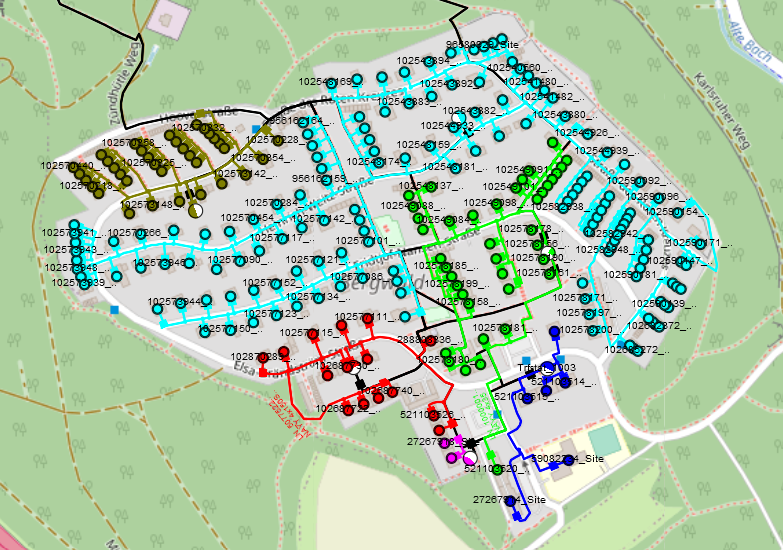}
    \caption{Case $(EM, T_{C})$: Grid model using the \ac{DSO} data on the number of electricity meters at each building. The number and location of transformers are calculated.}     
    \label{fig:em-tc}
\end{subfigure}

\caption{Distribution grid modeling with a \textit{calculated} number and position of substations using different data sources for the load modeling comprising the combinations $(X, T_{C})$. The locations of the transformers are calculated with a \textit{k-means} approach for the spatial load distribution in each case. The color of the nodes and lines indicates the affiliation to a specific transformer. As there is no one-to-one relation between the transformers of the different variants, the coloring is not consistent.}
\label{fig:2all}
\end{figure*}

\section{Evaluation}
\label{sec:eval}

In this section, we describe the study area and the evaluation criteria before presenting the results.
A meaningful comparison between the proposed method and the methods found in literature is infeasible due to several factors, such as a focus on different voltage levels \cite{Amme.2018,Medjroubi.2017}, different grid styles \cite{Mateo.2020}, and missing implementation details \cite{Grzanic.2019,Klabunde.2022,Abhilash.2021}.
Thus, we focus on the evaluation of the impact of various available data sources and the comparison with real grid topology data provided by \ac{DSO}.

For this evaluation, we consider the three data sources introduced in \Cref{sec:method} and generate the distribution grid model either with a priori knowledge of the \trafovoltages{20}{0.4} transformer locations together with the \qty{20}{\kV} grid topology $(T_K)$ or with a calculation of the transformer locations and the \qty{20}{\kV} network topology $(T_C)$. 
This altogether results in six combinations of $(d,t)$ tuples for data source and transformer data, as given below:
\begin{equation}\label{eq:combinations}
(d,t)=\{O_{\mathit{2D}}, O_{\mathit{3D}}, EM\} \times \{T_{K},T_{C}\}
\end{equation}
With this wide variety of data source combinations, we aim to investigate the impact of data source quality on the quality of the resulting grid models.
In addition to the six models generated by these combinations, we consider a topological model $(DSO)$ based on GIS data provided by the \ac{DSO} of the target area.
This model is not generated by the method described in this paper, and has two significant differences compared to the generated models:
First, it includes switching devices that allow the reconfiguration of the grid topology, and second, it contains two separate cables for most of the streets, one on each side.

For the analysis of the generated models, graph metrics are applied to the automatically generated topologies, and the results of network calculations based on voltage drops and line loadings are compared.
While the $(DSO)$ model is considered as the reference model for the topological comparison of the models, we consider the $(EM, T_{K})$ as a reference model for evaluation of the electrical properties.
This is because the $(DSO)$ model lacks information regarding the demand, whereas the $(EM, T_{K})$ is the generated model based on the most accurate available data, i.e., electricity meter data and known transformer positions.
From these comparisons, statements are made regarding the quality of the automatically generated networks.
In particular, an answer is given to the question which data are sufficient for an automated modeling of the power grid for reliable statements.

\subsection{Study Area}
The study area is selected as in the analysis in \cite{Cakmak.2022.ISGT} where detailed information was collected from an on-site inspection.
This allows a direct comparison of the generated grid topology with previous results. 
The study area spans over roughly half a square kilometer and contains \num{241} buildings, including single houses, duplex houses, town houses as well as apartment towers. 
Even though the area is mostly residential, it also encompasses non-residential buildings such as a school, a community center and some shops. 
As the few shops in the area are in buildings that also encompass residential units, for this analysis they are treated as residential units for the sake of simplicity. 
The consumption of buildings with a purely non-residential use is calculated using the overall area available in the building multiplied by the average \si{\kWh} per \si{\m\squared} for the building type. 
For the present school and kindergarten, this value is \num{20} and \qty{22}{\kWh} per \si{\m\squared} per year respectively, and for the community center it is \qty{9}{\kWh} per \si{\m\squared} per year\cite{Kluttig.2001, Kirche.2019}. 
For the overall area, the number of levels of a building is obtained either from the level tag within the \ac{OSM} dataset or assumed to be two, thus $S_U = 2$ in this case.

\subsection{Load Estimation Comparison}

Comparing the load estimations based on available 2D and 3D \ac{OSM} data to the reference estimation based on \ac{DSO} supplied \acf{EM} data results in the deviations shown in \Cref{tab:load_estimation}.
The table shows that the 2D-based estimations result in a significantly lower load overall, while still containing buildings with significantly higher load estimations.
Thus, a simple adjustment of the scaling factor $S_U$ would not be sufficient, as a higher value would also increase the positive deviations.
On the other hand, the 3D-based estimations, while also containing some buildings with large deviations, match the overall load estimate very closely and perform better than a scaled 2D-based estimate would.

\begin{table}
    \centering
    \caption{Deviation of \ac{OSM}-based load estimations compared to the load estimation based on \ac{DSO}-supplied electricity meter data.}
    \label{tab:load_estimation}
    \begin{tabular}{lrr}
        \toprule
         & $O_{\mathit{2D}}$ & $O_{\mathit{3D}}$ \\
         \midrule
         Total deviation & \qty{-58}{\percent} & \qty{-3}{\percent} \\
         Greatest negative building deviation & \qty{-95}{\percent} & \qty{-69}{\percent} \\
         Greatest positive building deviation & \qty{+65}{\percent} & \qty{+85}{\percent} \\
         \bottomrule
    \end{tabular}
\end{table}

\subsection{Topological Comparison}

In this section, we perform a topological comparison of the models generated using the different data sources and generation methods.
First, we describe our findings on the geographic representations of the models before quantizing them using two graph metrics, i.e., the number of nodes per transformer and the eccentricity of the transformers.

\paragraph{Geographic representation}

\begin{figure}
    \centering
    \includegraphics[width=0.9\linewidth]{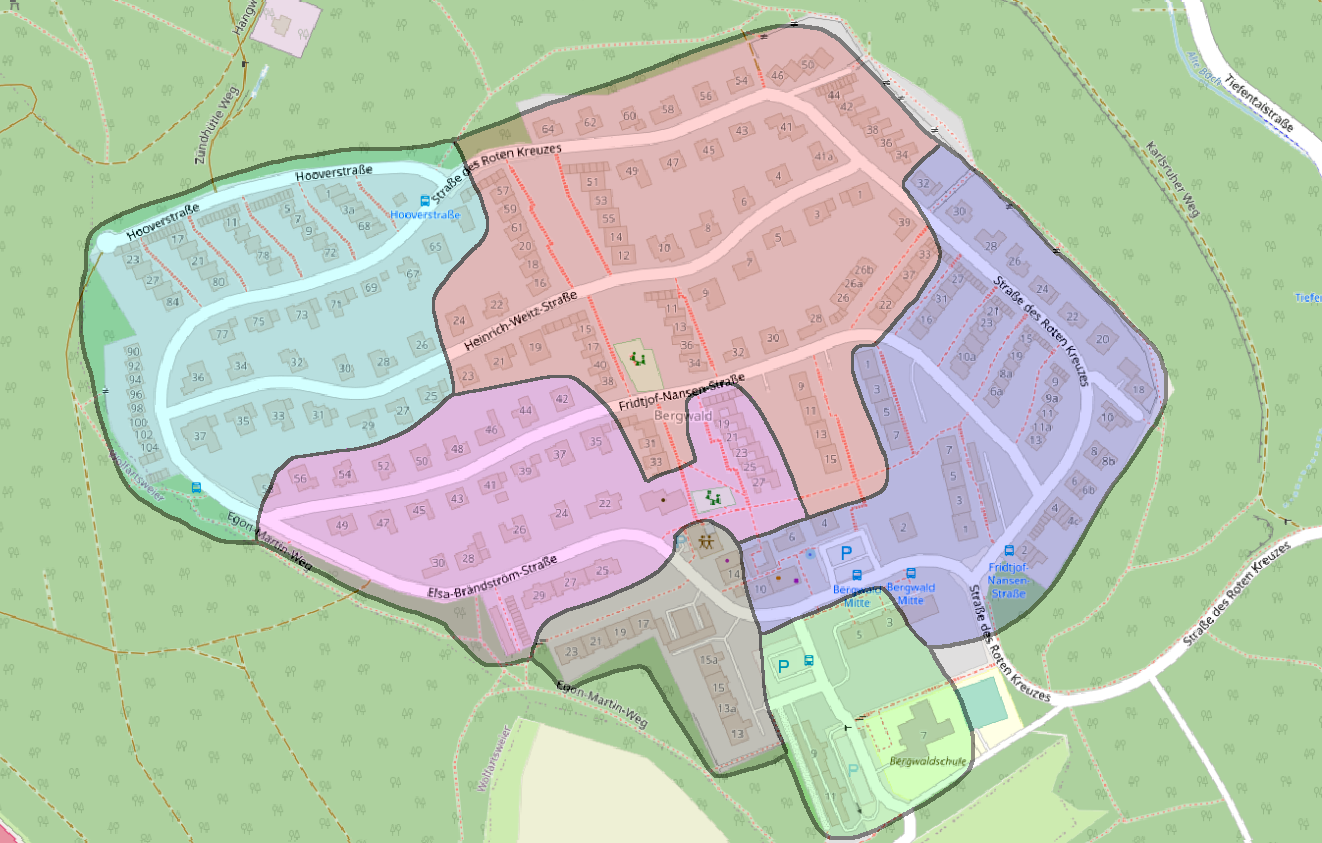}
    \caption{A simplified representation of the $(DSO)$ model, showing the areas supplied by the six different transformers. The coloring is equivalent to \Cref{fig:1all}.}
    \label{fig:dso-topology}
\end{figure}

The model in \Cref{fig:o2d-tk}, generated by $(O_{\mathit{2D}}, T_{K})$, shows six \qty{400}{\volt} subgrids with similar sizes that are not very compact and in some cases spread across the whole study area.
Especially notable are the two cables at the right edge of the area.
\Cref{fig:o3d-tk} shows the model created by $(O_{\mathit{3D}}, T_{K})$, which includes much more compact subgrids with larger size differences.
Notably, the brown and green subgrids are significantly smaller compared to the previous case.
\Cref{fig:em-tk} shows the $(EM, T_{K})$ model with even smaller subgrids in the lower part of the study area.
Consequently, the remaining subgrids are significantly larger.
This model also includes an unrealistic connection via a footpath in the brown subgrid.
Examining the models with calculated transformer positions, it is noticeable that the $(O_{\mathit{2D}}, T_{C})$ case, as shown in \Cref{fig:o2d-tc}, contains only two transformers compared to the usual six transformers in all other models.
The $(O_{\mathit{3D}}, T_{C})$ model in \Cref{fig:o3d-tc} presents a realistic partitioning into subgrids, with two smaller subgrids in the lower part and four evenly distributed larger subgrids in the upper part of the study area.
The $(EM, T_{C})$ in \Cref{fig:em-tc} contains very unevenly distributed subgrids with a high variance in size.
Compared to the six generated models, the $(DSO)$ model offers flexibility in its topology, as it contains switchgear at 15 different locations with a variety of switching options, resulting in a very high number of interconnection variants.
Since the data provided by the \ac{DSO} does not include details about the switching states of the switchgear, some assumptions are required to obtain a valid configuration.
The configuration that results in the partitioning shown in \Cref{fig:dso-topology} is based on the transformer positions and aims to balance the consumption in each area while keeping each area compact.
Overall, the $(DSO)$ model shows the closest similarity to the $(O_{\mathit{3D}}, T_{C})$ version of the generated models, which requires the least a priori knowledge of sensitive data.

\paragraph{Number of nodes per transformer}

In this metric, a node represents a busbar in the generated PowerFactory model.
For each building, our generation methods create two busbars: One for the building itself and one at the connection of the main cable and the house cable.
The number of nodes per transformer is calculated by determining the supplying transformer for each \qty{400}{\volt} busbar.
\Cref{fig:eval:nodes} shows the number of nodes per transformer for the six generated models.
The two models using only the 2D \ac{OSM} data reveal a very even distribution of nodes between the different transformers, due to the homogeneity of the estimated loads.
Furthermore, it is noteworthy that the $(O_{\mathit{2D}}, T_{C})$ case consists of only two transformers with around 250 connected nodes each, which is highly unrealistic. According to \ac{DSO} data, one \qty{630}{\kilo\voltampere} transformer, which is the installed transformer type in the study area, serves on average 79 residential units.
The other cases show a more heterogeneous distribution of nodes due to the large differences in building load estimations, that are far more realistic for some multi-story buildings.
While containing more nodes overall due to some implementation details, the $(DSO)$ model located in between the $(O_{\mathit{3D}}, T_{K})$ and $(EM, T_{K})$ variants with a good matching to $(O_{\mathit{3D}}, T_{C})$.

\paragraph{Eccentricity per transformer}

The eccentricity of a node is defined as the maximum distance to all other nodes in a graph.
Thus, when determining the eccentricity of a transformer $T_i$, we calculate
\begin{equation}\label{eq:eccentricity}
    ecc(T_i)=\max_{n \in N_{T_i}}(d(T_i,n)),
\end{equation}
where $N_{T_i}$ is the set of \qty{400}{\volt} nodes that are supplied by $T_i$, and $d$ is the cable distance between two nodes.
Essentially, the eccentricity of a transformer describes the maximum cable length between the transformer and the buildings supplied by it.
The comparison of eccentricities in the generated models is illustrated in \Cref{fig:eval:eccentricity}.
This plot reveals two outliers in the $(O_{\mathit{2D}}, T_{C})$ case with an eccentricity of \qty{2.1} and \qty{1.2}{\kilo\meter} respectively, while all other transformers have an eccentricity below \qty{0.75}{\kilo\meter}.
The $(DSO)$ model features very similar eccentricity values to the $(O_{\mathit{3D}}, T_{C})$ variant with a small range of values, because of the configuration criterion to keep the individual partitions compact.
In general, the $EM$-based models exhibit a wider distribution of eccentricities than their $O_{\mathit{3D}}$-based counterparts, whereas the $T_C$-based models have overall lower eccentricities than their $T_K$-based equivalents.

\begin{figure}
    \centering
    \begin{tikzpicture}
    \pgfplotsset{grid style={dotted}, every x tick label/.append style={font=\small, yshift=0.5ex}}
        \begin{axis}[
            height=5.5cm,
            width=0.99\linewidth,
            ytick={0,50,...,300}, 
            grid=major,
            ylabel={\# of Nodes per Transformer},
            symbolic x coords={2D-K,3D-K,EM-K,2D-C,3D-C,EM-C,DSO-K},
            xticklabels={0,$(O_{\mathit{2D}}{,}T_{K})$, $(O_{\mathit{3D}}{,}T_{K})$, $(EM{,}T_{K})$, $(O_{\mathit{2D}}{,}T_{C})$, $(O_{\mathit{3D}}{,}T_{C})$, $(EM{,}T_{C})$, $(DSO)$},
            x tick label style={rotate=30},
        ]
        \addplot [
        scatter,%
        scatter/@pre marker code/.code={%
            \edef\temp{\noexpand\definecolor{mapped color}{HTML}{\pgfplotspointmeta}}%
            \temp
            \scope[draw=mapped color!80!black,fill=mapped color]%
        },%
        scatter/@post marker code/.code={%
            \endscope
        },%
        only marks,     
        mark=*,
        fill opacity=0.6,
        point meta={TeX code symbolic={%
            \edef\pgfplotspointmeta{\thisrow{color}}%
        }},
    ]  table[col sep=semicolon, x index=0, y index=2]{DATA/trafo-data.csv};
        \end{axis}
    \end{tikzpicture}
    \caption{The number of nodes per transformer shows a very even distribution for both $O_{\mathit{2D}}$ cases and a wider distribution for the other cases.}
    \label{fig:eval:nodes}
\end{figure}
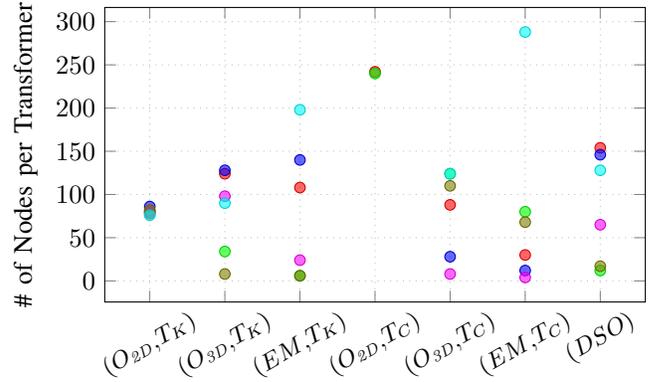

\begin{figure}
    \centering
    \begin{tikzpicture}
        \pgfplotsset{grid style={dotted}, every x tick label/.append style={font=\small, yshift=0.5ex}}
        \begin{axis}[
            height=5.5cm,
            width=0.99\linewidth,
            grid=major,
            ylabel={Eccentricity per Transformer [km]},
            symbolic x coords={2D-K,3D-K,EM-K,2D-C,3D-C,EM-C,DSO-K},
            xticklabels={0,$(O_{\mathit{2D}}{,}T_{K})$, $(O_{\mathit{3D}}{,}T_{K})$, $(EM{,}T_{K})$, $(O_{\mathit{2D}}{,}T_{C})$, $(O_{\mathit{3D}}{,}T_{C})$, $(EM{,}T_{C})$, $(DSO)$},
            x tick label style={rotate=30},
        ]
        \addplot [
        scatter,%
        scatter/@pre marker code/.code={%
            \edef\temp{\noexpand\definecolor{mapped color}{HTML}{\pgfplotspointmeta}}%
            \temp
            \scope[draw=mapped color!80!black,fill=mapped color]%
        },%
        scatter/@post marker code/.code={%
            \endscope
        },%
        only marks,     
        mark=*,
        fill opacity=0.6,
        point meta={TeX code symbolic={%
            \edef\pgfplotspointmeta{\thisrow{color}}%
        }},
        ] table[col sep=semicolon, x index=0, y index=5]{DATA/trafo-data.csv};
        \end{axis}
    \end{tikzpicture}
    \caption{The eccentricity per transformer shows the maximum distance between a transformer and the connected \qty{400}{\volt} nodes. This metric reveals two very large distances for the first case, while the other cases show more evenly distributed eccentricities. In general, the EM-based cases show a wider distribution than the 3D-data-based cases.}
    \label{fig:eval:eccentricity}
\end{figure}
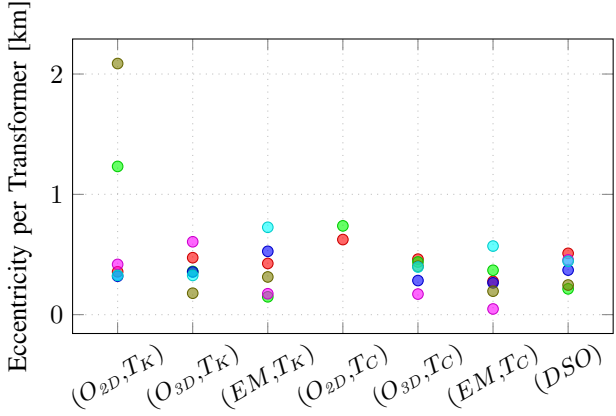

\subsection{Electrical Comparison}

In this section, we perform a comparison of the electrical properties of the generated distribution grid models.
In a first step, we analyze the voltage profiles of all cases and verify that the voltage profiles are within the valid voltage range.
In particular, we compare the voltage profiles of the reference model $(EM, T_{K})$ shown in \Cref{fig:voltage_em-tk} with the proposed method $(O_{\mathit{3D}}, T_{C})$ shown in \Cref{fig:voltage_o3d-tc}.
The proposed method based on public data and transformer estimation outperforms regarding the voltage range.
Both are in a valid range, but the case $(O_{\mathit{3D}}, T_{C})$ has a significantly narrower voltage band of $[0.989, 0.999]$ \si{\pu} compared to the reference model, with $[0.967, 0.998]$ \si{\pu} 
This also correlates with the shorter radial feeder length of \qty{0.461}{\km} compared to \qty{0.726}{\km} in the reference model.
In both diagrams, the voltage profile with the steepest drop is associated with a transformer with high loads in multi-story buildings supplied with short length cables.
Thus, this is another indicator of the appropriate choice of methodology for approximating the location and number of transformers, based on load estimation from building data, which itself also seems promising. 

\begin{figure}[t]
\includegraphics[width=\linewidth]{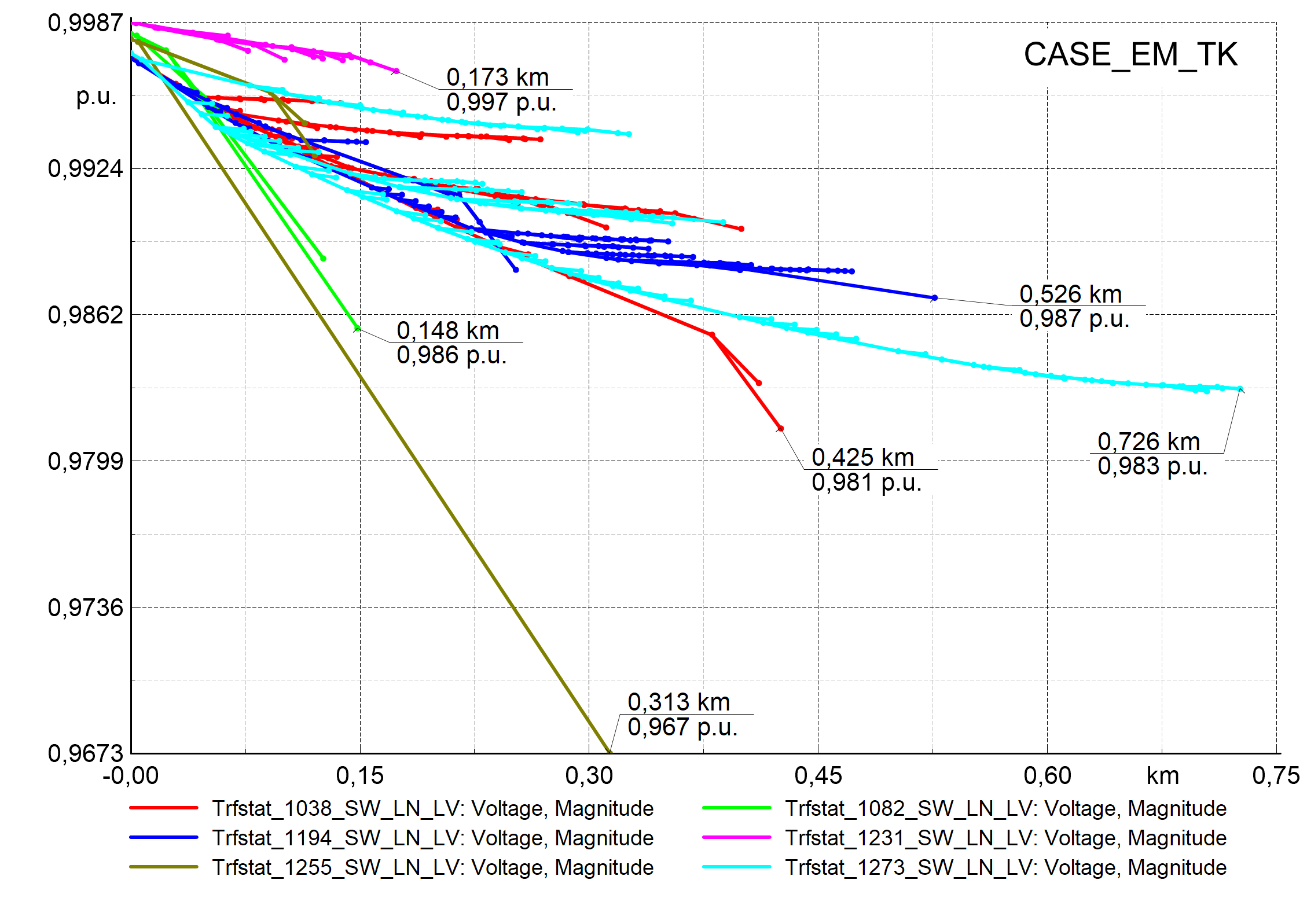}
    \caption{The voltage profiles for case $(EM, T_{K})$ show steep decreases for the brown and green feeders, which contain multifamily buildings with around 70 smart meters each.}
    \label{fig:voltage_em-tk}
\end{figure}
\begin{figure}[t]
    \includegraphics[width=\linewidth]{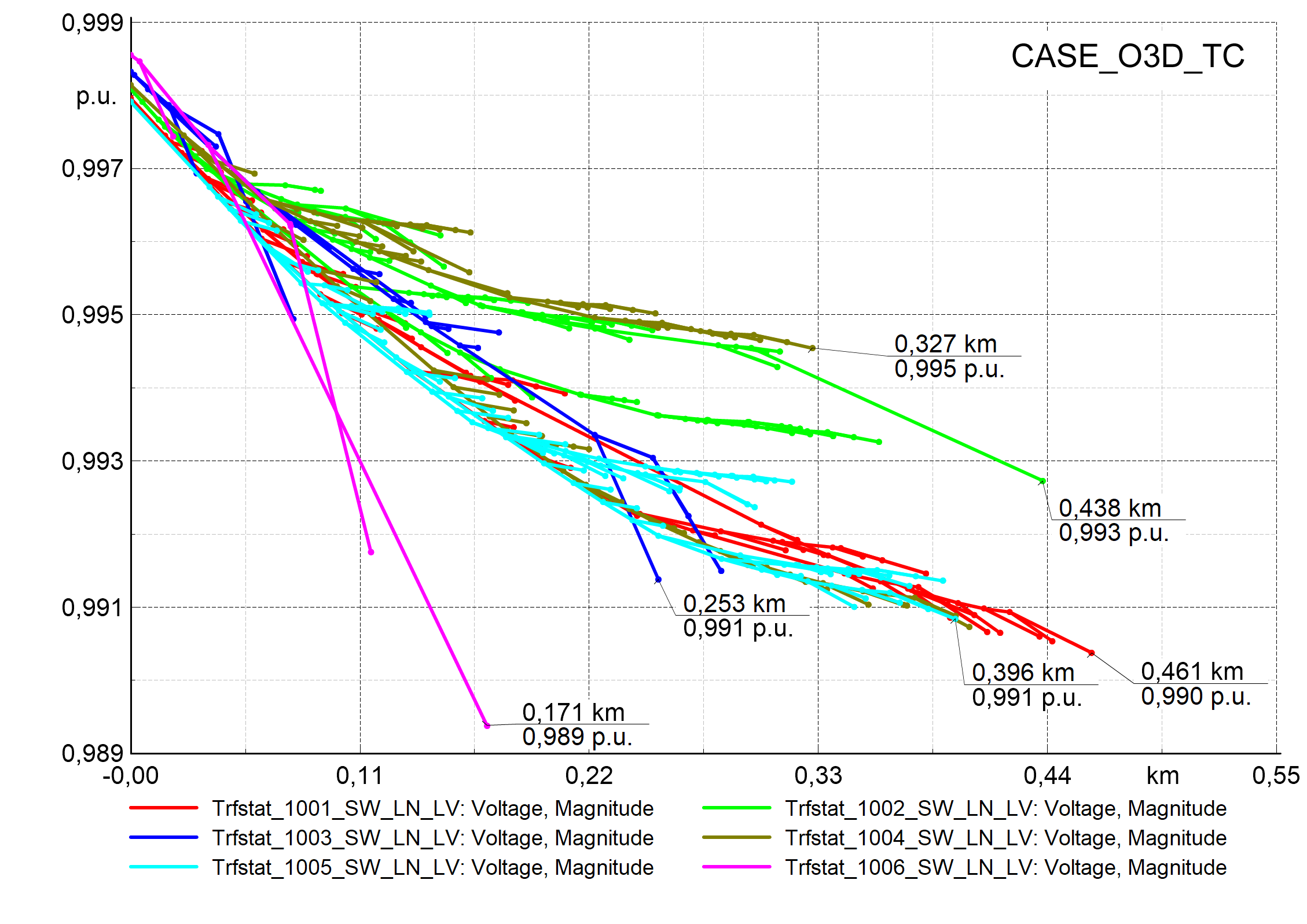}
    \caption{The voltage profiles for case $(O_{3D}, T_{C})$ also show a steep decline for the feeder containing the multifamily buildings (pink). In this case, however, the overall voltage band is much more narrow.}
    \label{fig:voltage_o3d-tc}
\end{figure}

Further, we analyze the statistical distribution of line loadings of the \qty{0.4}{\kV} cables as shown in \Cref{fig:histogram_lineloading}. 
First, it is noticeable that the distributions for the $(O_{\mathit{2D}}, X)$ cases differ seriously from the others.
This is due to the weak load estimation based on 2D \ac{OSM} data that leads to overall lower line loadings.
To further analyze the similarities of the line loading distributions, a similarity index is defined based on the Euclidean distance of the histogram distributions.
The full table of similarity indices in pairs of the cases is given in \Cref{tab:histo_dist}.
The highest similarity, i.e. smallest distance, with the reference case $(EM, T_{K})$ (shown in red in \Cref{tab:histo_dist}) is observable for case $(EM, T_{C})$ with $15.68$, where both share the same smart meter data distribution over the buildings.
The second-best match with the reference is the case $(O_{\mathit{3D}}, T_{K})$ with $57.48$, where the transformer number and positions are identical and a priori known.
The very close third-best option is $(O_{\mathit{3D}}, T_{C})$ with $59.7$, which does not need any additional input data for network topology generation at all and seems a promising approach for future use.
The second finding from the similarity index table concerns the quality of the approximation of the number and location of the transformer stations. 
For this purpose, we compare given ($T_K$) and calculated ($T_C$) transformer properties for each case.
The values shown in blue in \Cref{tab:histo_dist} indicate a high accordance between the ($T_C$) and corresponding ($T_K$) models.

Note that this is only a statistical comparison without consideration of the real spatial distribution, which is handled in \Cref{chap:GIS-comparison}.

\begin{figure}[t]
    \includegraphics[width=\linewidth]{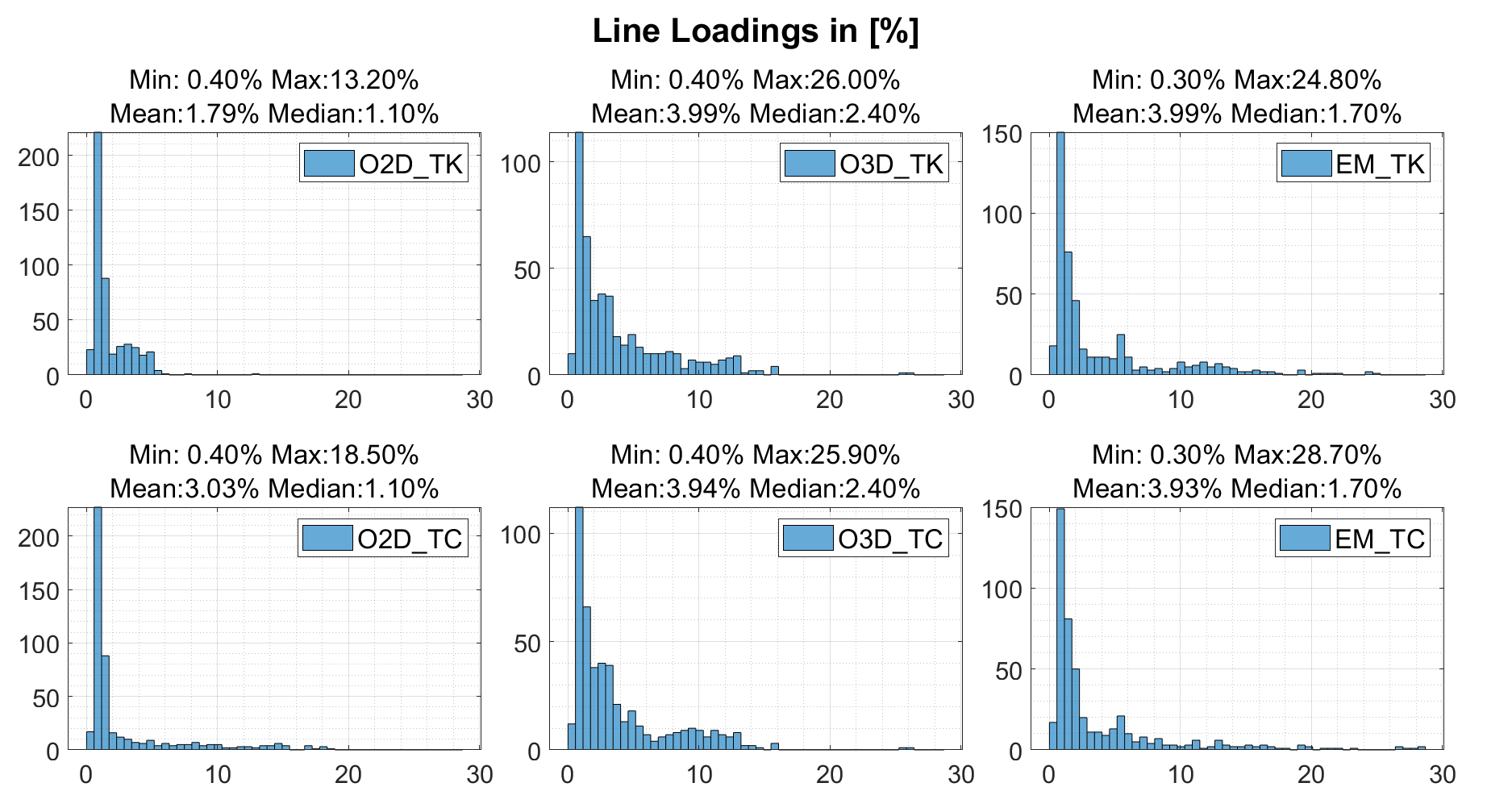}
    \caption{The histograms of line loadings at \qty{0.4}{\kV} show a high accordance between the $O_{\mathit{3D}}$ and $EM$-based models. Furthermore, the line loadings in these models are only slightly affected by the transformer placement method.}
    \label{fig:histogram_lineloading}
\end{figure}

\begin{figure}[th]
    \includegraphics[width=\linewidth]{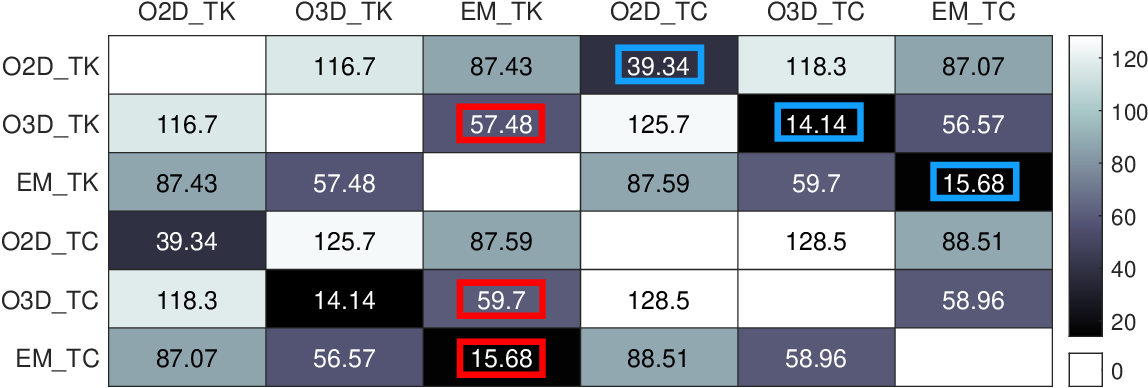} 
    \caption{The similarity indices of the cases based on histograms of line loadings show a high accordance between models with the same load estimation and different transformer placements (smaller values indicate high similarity). Furthermore, they confirm the similarity of the $O_{\mathit{3D}}$ and $EM$ models.}
    \label{tab:histo_dist}
\end{figure}

\subsection{GIS-based Comparison}
\label{chap:GIS-comparison}
To be able to evaluate the approach for the automatic \qty{20}{\kV} network generation, the calculated transformer positions $T_{C}$ are compared to the known positions $T_{K}$ for selected cases. 
This includes the comparison of cases $(EM, T_{C})$ and $(O_{\mathit{3D}}, T_{C})$ to case $(EM, T_{K})$.
In particular, each calculated position is mapped to the closest known transformer.
If this mapping is not unambiguous, the closest calculated position is chosen to be mapped to the known transformer location in question.
That way, there is a one-to-one comparison, the results of which are depicted in \Cref{tab:trafo_distance}.

Given that the study area measures \qty{582}{\meter} from the most northern to most southern point and \qty{766}{\meter} from east to west, the rather low distances mean that the methodology is working well. 
From the spatial vicinity of the calculated and known transformers, it becomes evident that the cluster-based computation for the transformer locations yields good results.
On the other hand, the fact that the values change not much from the $EM$ to the $O_{\mathit{3D}}$ case means that the \ac{OSM} data with height information of buildings is sufficient for good results for transformer placement. 

\begin{table}[h]
    \centering
    \caption{The results of the location-based comparison show very similar distances between known and calculated transformer positions for the $O_{\mathit{3D}}$ and $EM$ model.}
    \label{tab:trafo_distance}{
        \begin{tabular}{*5{c}}  
            \toprule
            & \multicolumn{2}{c}{Distances from case $(DSO)$} \\
            Transformer No. & to $(O_{\mathit{3D}}, T_{C})$ & to $(EM, T_{C})$\\
            \midrule
            \small{1}  & \qty{150}{\meter}    & \qty{148}{\meter}\\
            \small{2}  & \qty{83}{\meter}    & \qty{72}{\meter}\\
            \small{3}  & \qty{37}{\meter}     & \qty{42}{\meter}\\
            \small{4}  & \qty{116}{\meter}    & \qty{105}{\meter}\\
            \small{5}  & \qty{50}{\meter}    & \qty{37}{\meter}\\
            \small{6}  & \qty{127}{\meter}   & \qty{123}{\meter}\\
            \bottomrule
        \end{tabular}
        }
        \vspace{-1mm}
\end{table}

\subsection{Runtime Evaluation}

The runtime of the introduced Inverse Proportional Cable Capacity Estimation (IP-CCE) method \eqref{eq:cap} as part of the optimization method in the course of low voltage grid topology generation (see \Cref{sec:method:stage2}) is evaluated with a comparison to the Newton bisection method in \cite{Cakmak.2022.ISGT}.

To obtain the runtimes, the application was timed on a Windows machine with 32 GB of RAM storage and an Intel\textregistered\ Core\texttrademark\ i7-10700K CPU with 8 physical cores running at 3.8 GHz.
As can be seen in \Cref{fig:runtime}, the runtime of the optimization was significantly reduced by \num{1307} seconds, reducing the total runtime of the model generation from \num{1618} to \num{311} seconds.
This results in an overall speedup factor of \num{5.2}.
While most of the programs steps do not differ and therefore stay roughly the same, the difference in the optimization process can be clearly seen.
This improvement is especially important because with bigger models to be created, the optimization time is expected to grow exponentially.

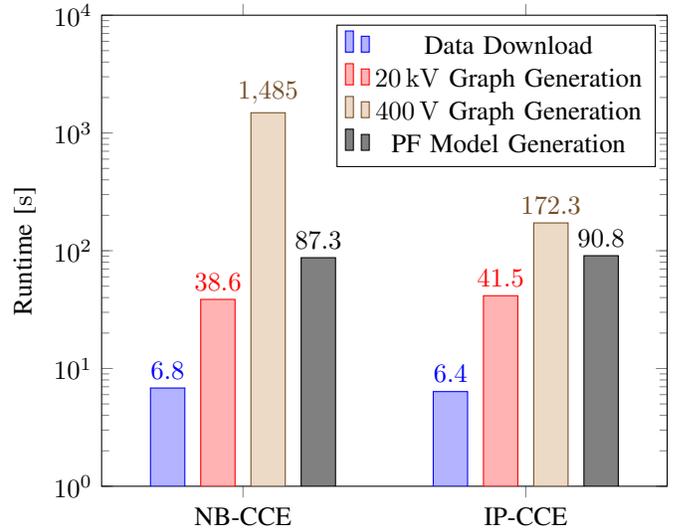
\begin{figure}[t]
    \begin{tikzpicture}
        \begin{semilogyaxis}[
            ybar=6pt,
            ylabel={Runtime [\unit{\second}]},
            width=9.1cm,
            symbolic x coords={NB-CCE,IP-CCE},
            xtick=data,
            enlarge x limits=0.5,
            ymin=1,
            ymax=10^4,
            nodes near coords,
            point meta=rawy,
            bar width=13pt,
            nodes near coords style={/pgf/number format/.cd,precision=1}
        ]
        \addplot table[col sep=semicolon, x index=0, y index=1]{DATA/runtime-aggregated-final.csv};
        \addplot table[col sep=semicolon, x index=0, y index=2]{DATA/runtime-aggregated-final.csv};
        \addplot table[col sep=semicolon, x index=0, y index=3]{DATA/runtime-aggregated-final.csv};
        \addplot table[col sep=semicolon, x index=0, y index=4]{DATA/runtime-aggregated-final.csv};
        \legend{Data Download, \qty{20}{\kilo\volt} Graph Generation, \qty{400}{\volt} Graph Generation, PF Model Generation}

        \end{semilogyaxis}
    \end{tikzpicture}
\caption{Runtime comparison of the introduced Inverse Proportional estimation (IP-CCE) to the Newton Bisection method (NB-CCE) from \cite{Cakmak.2022.ISGT} in seconds. A speedup by a factor of \num{8.6} is achieved for the optimization with the new method for cable capacity estimation. The \qty{20}{\kilo\volt} graph generation includes the steps described in \Cref{sec:method:stage1}, while the \qty{400}{\volt} graph generation comprises \Cref{sec:method:stage2}. The last step includes the generation of a PowerFactory model as described in \Cref{sec:method:powerfactory}.}
\label{fig:runtime}
\end{figure}

\section{Discussion}
\label{sec:discussion}

The presented new model generation approach enables the automated generation of distribution grid models from minimal open data sources.
The evaluation of the models generated using various data sources, from open 2D and 3D data to proprietary electricity meter data and GIS data including transformer positions, demonstrates the viability of open data sources for the automated generation of realistic distribution grid models.
However, the evaluation also shows that the 2D data alone, as available in \ac{OSM}, does not yield realistic models with the presented approach.
On the other hand, combined with the height information found in OSM Buildings, the generated models closely resemble the models utilizing proprietary electricity meter data supplied by a local \ac{DSO}, which confirms the viability of the household estimation method.

While the load estimation is shown to be crucial for the subsequent estimation of the transformer number and positions and the cabling, the transformer placement itself is less important for the evaluated metrics.
Especially for the loadings of the \qty{400}{\volt} lines, the transformer placement is nearly negligible, as \Cref{tab:histo_dist} shows.
In general, however, the transformers placed by the deployed optimization algorithm lead to more efficient cable layouts (see \Cref{fig:eval:eccentricity}).

Overall, the evaluation shows the viability of our new method to generate realistic distribution grid models from openly available building data.
In some instances, the estimated household numbers might even be more accurate than the \ac{DSO} data, since some buildings contain an unrealistic high number of meters.
However, a thorough comparison to a real model under various load scenarios still needs to be performed.
Furthermore, the load estimation method proposed in the present work is limited to residential buildings and needs to be expanded to nonresidential and mixed buildings.

\section{Conclusion and Outlook}
\label{sec:conclusion}

The present paper introduces a new distribution grid model generation method, that enables the automated generation of grid models relying solely on openly available data sources, i.e., \acf{OSM} and OSM Buildings.
It introduces a building load estimation method that is based on the estimation of households that utilizes open 2D and 3D building data.
While the household estimation based on 2D data results in a poor model quality, our evaluation shows that the 3D-based estimation performs similarly to proprietary electricity meter data supplied by the \acf{DSO} of the study area.
While our evaluation highlights the importance of the available data for the load estimation, it also shows the relatively small importance of the actual \trafovoltages{20}{0.4} transformer placement for the generation of realistic models.

In the future,  we will consider seasonal \ac{PV} generation and thus residual loads together with further methods for the load estimation.
These include a way of incorporating commercial and industrial areas and buildings into the model, as well as a way to automatically detect the kind of residential areas that are present in order to be able to individually handle them. 
Furthermore, our approach has to be further developed in order to be able to accommodate more varying and challenging network topologies such as inner-city blocks with buildings inside the inner courtyard or overhead lines that are still present around our study area and thus in Germany.
Moreover, as seen in the $(DSO)$ model, real grids contain switchgear that allows the reconfiguration of the network topology to react to certain grid events.
We will work on methods to incorporate these switching capabilities into the model generation process, and to find realistic configurations for the switchgear.
Additionally, the comparison of various data sources and methods will help to select the best suitable open data source option for automated parallelized large-scale modeling of the power system by embedding the distribution grid models into the higher voltage levels and finally to co-simulation with other energy sectors.

\bibliographystyle{IEEEtran} 
\bibliography{bibliography.bib} 

\begin{thebibliography}{10}
\providecommand{\url}[1]{#1}
\csname url@samestyle\endcsname
\providecommand{\newblock}{\relax}
\providecommand{\bibinfo}[2]{#2}
\providecommand{\BIBentrySTDinterwordspacing}{\spaceskip=0pt\relax}
\providecommand{\BIBentryALTinterwordstretchfactor}{4}
\providecommand{\BIBentryALTinterwordspacing}{\spaceskip=\fontdimen2\font plus
\BIBentryALTinterwordstretchfactor\fontdimen3\font minus
  \fontdimen4\font\relax}
\providecommand{\BIBforeignlanguage}[2]{{%
\expandafter\ifx\csname l@#1\endcsname\relax
\typeout{** WARNING: IEEEtran.bst: No hyphenation pattern has been}%
\typeout{** loaded for the language `#1'. Using the pattern for}%
\typeout{** the default language instead.}%
\else
\language=\csname l@#1\endcsname
\fi
#2}}
\providecommand{\BIBdecl}{\relax}
\BIBdecl

\bibitem{Zinaman.2018}
\BIBentryALTinterwordspacing
O.~Zinaman, S.~Mueller, P.~Vithayasrichareon, and E.~Gutierrez, ``Status of
  power system transformation 2018: Advanced power plant flexibility,'' 2018,
  accessed: 2022-01-30. [Online]. Available:
  \url{https://www.21stcenturypower.org/assets/pdfs/main-report.pdf}
\BIBentrySTDinterwordspacing

\bibitem{Miller.2015}
\BIBentryALTinterwordspacing
{M. Miller et al.}, ``Status report on power system transformation: A 21st
  century power partnership report,'' {U.S. Department of Energy}, Tech. Rep.,
  2015, accessed: 2022-01-30. [Online]. Available:
  \url{https://www.nrel.gov/docs/fy15osti/63366.pdf}
\BIBentrySTDinterwordspacing

\bibitem{Cakmak.2022.ISGT}
H.~K. Çakmak, L.~Janecke, M.~Weber, and V.~Hagenmeyer, ``An optimization-based
  approach for automated generation of residential low-voltage grid models
  using open data and open source software,'' in \emph{2022 IEEE PES Innovative
  Smart Grid Technologies Conference Europe (ISGT-Europe)}, 2022, pp. 1--6.

\bibitem{OSMContrib.2004}
\BIBentryALTinterwordspacing
{OpenStreetMap Contrib.}, ``Openstreetmap,'' 2004, accessed: 2022-01-30.
  [Online]. Available: \url{https://www.openstreetmap.org/}
\BIBentrySTDinterwordspacing

\bibitem{OSMBuildings.2022}
\BIBentryALTinterwordspacing
{OSMBuildings Developers}, ``Osmbuildings,'' accessed: 2022-12-22. [Online].
  Available: \url{https://github.com/OSMBuildings/OSMBuildings}
\BIBentrySTDinterwordspacing

\bibitem{MateoDomingo.2011.full}
C.~{Mateo Domingo}, T.~{Gomez San Roman}, A.~Sanchez-Miralles, J.~P. {Peco
  Gonzalez}, and A.~{Candela Martinez}, ``A reference network model for
  large-scale distribution planning with automatic street map generation,''
  \emph{IEEE Transactions on Power Systems}, vol.~26, no.~1, pp. 190--197,
  2011.

\bibitem{InstituteforEnergyandTransportJointResearchCentre2016DistributionSystem}
{Institute for Energy {and} Transport (Joint Research Centre)}, A.~Lucas,
  G.~Prettico, A.~Mengolini, G.~Fulli, and F.~Gangale, \emph{Distribution
  System Operators Observatory: From European Electricity Distribution Systems
  to Reference Network}.\hskip 1em plus 0.5em minus 0.4em\relax LU:
  Publications Office of the European Union, 2016.

\bibitem{Grzanic.2019}
\BIBentryALTinterwordspacing
M.~Grzanic, M.~G. Flammini, and G.~Prettico, ``Distribution network model
  platform: A first case study,'' \emph{Energies}, vol.~12, no.~21, 2019.
  [Online]. Available: \url{https://www.mdpi.com/1996-1073/12/21/4079}
\BIBentrySTDinterwordspacing

\bibitem{Mateo.2020}
C.~Mateo, F.~Postigo, F.~de~Cuadra, T.~G.~S. Roman, T.~Elgindy, P.~Dueñas,
  B.-M. Hodge, V.~Krishnan, and B.~Palmintier, ``Building large-scale u.s.
  synthetic electric distribution system models,'' \emph{IEEE Transactions on
  Smart Grid}, vol.~11, no.~6, pp. 5301--5313, 2020.

\bibitem{Krishnan.2020}
V.~Krishnan, B.~Bugbee, T.~Elgindy, C.~Mateo, P.~Duenas, F.~Postigo, J.-S.
  Lacroix, T.~G.~S. Roman, and B.~Palmintier, ``Validation of synthetic u.s.
  electric power distribution system data sets,'' \emph{IEEE Transactions on
  Smart Grid}, vol.~11, no.~5, pp. 4477--4489, 2020.

\bibitem{Palmintier.2021}
B.~Palmintier, T.~Elgindy, C.~Mateo, F.~Postigo, T.~Gómez, F.~{de Cuadra}, and
  P.~D. Martinez, ``Experiences developing large-scale synthetic u.s.-style
  distribution test systems,'' \emph{Electric Power Systems Research}, vol.
  190, p. 106665, 2021.

\bibitem{Abhilash.2021}
A.~Bandam, C.~Syranidou, J.~Linssen, and D.~Stolten, ``Geo-referenced synthetic
  low-voltage distribution networks: A data-driven approach,'' in \emph{2021
  IEEE PES Innovative Smart Grid Technologies Europe (ISGT Europe)}, 2021, pp.
  1--6.

\bibitem{Medjroubi.2017}
W.~Medjroubi, U.~P. M{\"u}ller, M.~Scharf, C.~Matke, and D.~Kleinhans, ``Open
  data in power grid modelling: New approaches towards transparent grid
  models,'' \emph{Energy Reports}, vol.~3, pp. 14--21, 2017.

\bibitem{Amme.2018}
J.~Amme, G.~Ple{\ss}mann, J.~B{\"u}hler, L.~H{\"u}lk, E.~K{\"o}tter, and
  P.~Schwaegerl, ``The ego grid model: An open-source and open-data based
  synthetic medium-voltage grid model for distribution power supply systems,''
  \emph{Journal of Physics: Conference Series}, vol. 977, no.~1, p. 012007,
  Feb. 2018.

\bibitem{Klabunde.2022}
F.~Klabunde, C.~Reinhold, and B.~Engel, \emph{{Regionsabh\"angige
  Energiesystemanalysen auf Basis einer datengesteuerten
  Verteilnetzmodellierung}}, Feb. 2022.

\bibitem{Matpower.2011}
R.~D. Zimmerman, C.~E. Murillo-Sánchez, and R.~J. Thomas, ``Matpower:
  Steady-state operations, planning, and analysis tools for power systems
  research and education,'' \emph{IEEE Transactions on Power Systems}, vol.~26,
  no.~1, pp. 12--19, 2011.

\bibitem{OpenDSS.2017}
D.~Montenegro, R.~C. Dugan, and M.~J. Reno, ``Open source tools for high
  performance quasi-static-time-series simulation using parallel processing,''
  in \emph{2017 IEEE 44th Photovoltaic Specialist Conference (PVSC)}, 2017, pp.
  3055--3060.

\bibitem{Cakmak.2022}
H.~K. Çakmak and V.~Hagenmeyer, ``Using open data for modeling and simulation
  of the {A}ll {E}lectrical {S}ociety in {eASiMOV},'' in \emph{2022 Open Source
  Modelling and Simulation of Energy Systems (OSMSES)}, 2022, pp. 1--6.

\bibitem{Meier.1999}
\BIBentryALTinterwordspacing
H.~Meier, C.~F{\"u}nfgeld, T.~Adam, and B.~Schieferdecker, ``Repr{\"a}sentative
  {VDEW} {L}astprofile,'' accessed: 2022-01-30. [Online]. Available:
  \url{bdew.de/media/documents/1999\_Repraesentative-VDEW-Lastprofile.pdf}
\BIBentrySTDinterwordspacing

\bibitem{Zensus.2011.dt}
\BIBentryALTinterwordspacing
{Statistische Ämter des Bundes und der Länder}, ``Zensus 2011,'' accessed:
  2022-12-22. [Online]. Available: \url{https://www.zensus2011.de}
\BIBentrySTDinterwordspacing

\bibitem{bde.2004}
\BIBentryALTinterwordspacing
B.~der Energieverbraucher. (2004) Die stromformel für verbraucher. Accessed:
  2023-05-30. [Online]. Available:
  \url{https://www.energieverbraucher.de/de/bewertung-des-stromverbrauchs__646/ContentDetail__3449/}
\BIBentrySTDinterwordspacing

\bibitem{BW_GebEnergie_Gesetz.2020}
\BIBentryALTinterwordspacing
{Landesrecht BW}, ``{Gesetz zur Einsparung von Energie und zur Nutzung
  erneuerbarer Energien zur Wärme- und Kälteerzeugung in Gebäuden,
  Gebäudeenergiegesetz Anlage 5 (zu § 31 Absatz 1): Vereinfachtes
  Nachweisverfahren für ein zu errichtendes Wohngebäude},'' accessed:
  2022-12-22. [Online]. Available: \url{https://www.landesrecht-bw.de}
\BIBentrySTDinterwordspacing

\bibitem{Agricola.2012}
\BIBentryALTinterwordspacing
A.-C. Agricola, B.~H{\"o}flich, P.~Richard, J.~V{\"o}lker, C.~Rehtanz,
  M.~Greve, B.~Gwisdorf, J.~Kays, T.~Noll, J.~Schwippe, A.~Seack, J.~Teuwsen,
  G.~Brunekreeft, R.~Meyer, and V.~Liebert, ``{dena-Verteilnetzstudie: Ausbau-
  und Innovationsbedarf der Stromverteilnetze in Deutschland bis 2030},'' 2012,
  accessed: 2022-01-30. [Online]. Available:
  \url{https://www.dena.de/fileadmin/dena/Dokumente/Pdf/9100\_dena-Verteilnetzstudie\_Abschlussbericht.pdf}
\BIBentrySTDinterwordspacing

\bibitem{Lloyd.1982}
S.~Lloyd, ``Least squares quantization in pcm,'' \emph{IEEE Transactions on
  Information Theory}, vol.~28, no.~2, pp. 129--137, Mar. 1982.

\bibitem{dena.2012}
\BIBentryALTinterwordspacing
D.~E.-A.~G. (dena), ``dena-verteilnetzstudie. ausbau- und innovationsbedarf der
  stromverteilnetze in deutschland bis 2030.'' 2012, accessed: 2022-01-30.
  [Online]. Available:
  \url{https://www.dena.de/fileadmin/dena/Dokumente/Pdf/9100_dena-Verteilnetzstudie_Abschlussbericht.pdf}
\BIBentrySTDinterwordspacing

\bibitem{Stein.2022}
S.~Nickel, S.~Rebennack, O.~Stein, and K.-H. Waldmann, \emph{Operations
  Research}, 3rd~ed.\hskip 1em plus 0.5em minus 0.4em\relax Springer Gabler
  Berlin, 2022.

\bibitem{Christofides.1976}
N.~Christofides, ``Worst-case analysis of a new heuristic for the travelling
  salesman problem,'' Carnegie-Mellon Univ Pittsburgh Pa Management Sciences
  Research Group, Tech. Rep., 1976.

\bibitem{Kluttig.2001}
H.~Kluttig, A.~Dirscherl, and H.~Erhorn, ``{Energieverbr{\"a}uche von
  Bildungsgeb{\"a}uden in Deutschland},'' \emph{gi-Gesundheits-Ingenieur},
  no.~5, pp. 221--233, 2001.

\bibitem{Kirche.2019}
\BIBentryALTinterwordspacing
{Institut für Kirche und Gesellschaft der Ev. Landeskirche von Westfalen},
  ``{Energie-Bericht Ev.-luth. Kirchengemeinde Vahrenwald für das Jahr
  2019},'' 2019. [Online]. Available:
  \url{http://vahrenwalder-kirche.de/images/stories/kirche/02_Aktuelles/Energiebericht2019_ET.pdf}
\BIBentrySTDinterwordspacing

\end{thebibliography}

\end{document}